\documentclass{iaus}
\usepackage{graphicx}

\title[Reconstruction of the EUV irradiance] %% give here short title %%
{Towards the reconstruction of the EUV irradiance for solar cycle 23}

\author[Haberreiter]   %% give here short author list %%
{Margit Haberreiter
}

\affiliation{Physikalisch-Meteorologisches Observatorium Davos/World Radiation Center \\ 7260 Davos Dorf, Switzerland \\ email: {\tt margit.haberreiter@pmodwrc.ch}}

\author[Margit Haberreiter]   %% give here short author list %%
{Margit Haberreiter
 }

\affiliation{Physikalisch-Meteorologisches Observatorium Davos/World Radiation Center \\ 7260 Davos Dorf, Switzerland \\ email: {\tt margit.haberreiter@pmodwrc.ch} \\[\affilskip]
}
\pubyear{xxx}
\volume{xxx}  %% insert here IAU Symposium No.
\pagerange{xxx}
\jname{Title of your IAU Symposium}
\editors{A.C. Editor, B.D. Editor \& C.E. Editor, eds.}
\begin{document}

\maketitle

\begin{abstract}
We present preliminary reconstructions of the EUV from 26 to 34\,nm from February 1997 to May 2005, covering most of solar cycle 23. The reconstruction is based on synthetic EUV spectra calculated with the spectral synthesis code Solar Modeling in 3D (SolMod3D). These spectra are weighted by the relative area coverage of the coronal features as identified from EIT images. The calculations are based on one-dimensional atmospheric structures that represent a temporal and spatial mean of the chromosphere, transition region, and corona. The employed segmentation analysis considers coronal holes, the quiet corona, and active regions identified on the solar disk. The reconstructed EUV irradiance shows a good agreement with observations taken with the CELIAS/SEM instrument onboard SOHO. Further improvement of the reconstruction including more solar features as well as the off-limb detection of activity features will be addressed in the near future.

\end{abstract}

\firstsection % if your document starts with a section,
\section{Introduction}
The solar EUV irradiance is the main energy input for the upper Earth's atmosphere with important effects on the ionosphere and thermosphere. The solar energy output changes on short time-scales of minutes to hours as well as longer times-scales such as the 27-day solar rotation cycle or the 11-year solar cycle. There is also indication that the EUV irradiance might shows a secular trend (\cite[{Didkovsky} {et~al.} 2010]{Didkovsky2010ASPC}).

The EUV radiation incident on the upper Earth's atmosphere leads to a change in its temperature and density (see e.g. \cite[{Solomon} {et~al.} 2010]{Solomon2010}). In order to understand the effects of the changing EUV radiation on the Earth's atmosphere a continuous data set covering the short-term and long-term variations is essential. However, as space instruments are limited with regard to their temporal and spectral coverage, reliable models are needed to fill the gaps of the observational data sets.

  %\cite{FullerRowell2004,Lilensten2008AnGeo} it is clear that models that predict the thermospheric density require an improved knowledge of the incident EUV radiation.

{Several reconstruction approaches involve the use of proxies to describe the EUV variability. \cite[{Lean} {et~al.} (2011)]{Lean2011JGR} employ the two and three component NRLSSI model based on the Mg\,II and F$_{10.7}$ index to characterize and forecast the EUV variations. A further example for an empirical model is SOLAR2000 (\cite[{Tobiska} {et~al.} 2000]{Tobiska2000}), a model based on an extensive number of irradiance proxies. 

There is also ongoing work to determine which proxies, or spectral lines, are the best representatives for the variations of the entire EUV spectrum (see e.g. \cite[{Kretzschmar} {et~al.} 2009]{Kretzschmar2009AcGeo}, \cite[{Dudok de Wit} {et~al.} 2009]{DudokdeWit2009GRL}). 

Proxy models have been quite successful in describing the EUV variations, however, in order to understand the complete physical processes driving the irradiance variations, it is important to model the variations of the full solar spectra. \cite[Warren (2006)]{Warren2006} utilizes differential emission measure distributions derived from spatially and spectrally resolved solar observations and full-disk solar images. The reconstruction presented here follows the same principle as used by \cite[{Haberreiter} {et~al.} (2005)]{Haberreiter2005ASpR} and \cite[{Shapiro} {et~al.} (2011)]{Shapiro2011AA}. It includes the calculation of synthetic spectra with a radiative transfer code for various activity features on the solar disk. Weighting the spectra by filling factors derived from the relative area coverage of these activity features or from proxy data then yields the time dependent irradiance spectrum. 

In the following section the spectral synthesis code Solar Modeling in 3D (SolMod3D) is introduced. Then, in Section\,\ref{sec:rec} the reconstruction approach is described briefly. Finally, in Section\,\ref{sec:results} our results are compared with the EUV irradiance observations carried out with the CELIAS/SEM instrument (\cite[{Hovestadt} {et~al.}  1995]{SEM}) onboard the SOHO mission.

\section{Spectral Synthesis}{\label{sec:spectra}}
The spectral synthesis of the EUV is carried out with the SolMod3D code. It is a state-of-the-art radiative transfer code in full non-local thermodynamic equilibrium (NLTE). SolMod3D allows for the spherical line-of-sight integration which is very important for the correct calculation of the coronal emission. The code has already been successfully employed for the calculation of the quiet Sun EUV spectrum (\cite[Haberreiter 2011]{HaberreiterSoPh2011}) and for solar limb studies (\cite[Thuillier et al. 2011]{{Thuillier2011}}). 
 
The final goal is to reconstruct the EUV from 10 to 100\,nm. Here, however, we focus on the wavelength range from 26 to 34\,nm as observed with the SOHO/SEM instrument onboard SOHO. The calculations of the spectra are based on semi-em\-pi\-rical structures that represent a temporal and spatial mean of the photosphere, chromosphere, transition region and corona. For the photosphere, chromosphere, and transition region the full NLTE radiative transfer is solved based on the latest semi-em\-pi\-rical atmospheric structures by \cite[Fontenla et al. (2009)]{Fontenla2009ApJ}. The coronal lines are calculated as optically thin lines. Currently we employ four coronal structures; three represent different features of the quiet Sun, i.e. the quiet corona, coronal network and active coronal network as described in \cite[Haberreiter (2011)]{HaberreiterSoPh2011}. For the bright coronal regions a fourth structure is used. Details of this structure will be presented in a future publication. The coronal holes are currently represented with a contrast of 0.5 with respect to the quiet corona. 

It is important to note that these atmosphere structures constitute a temporal and spatial mean. As such, these structures fail to reproduce the very rich short-term variations of the solar atmosphere, in particular the corona. However, as we are mainly interested in the daily EUV variations, 1D structures are considered to be a suitable representation of the solar atmosphere. The advantage of the 1D atmosphere structures is that it allows us to take into account an extensive atomic data set (14,000 atomic levels and 170,000 spectral lines), essential for a realistic calculation of the EUV spectrum.

\section{Reconstruction}{\label{sec:rec}}
First, intensity spectra are calculated for different features on the solar disk as described above. These spectra are then weighted by filling factors derived from the analysis of solar images. For the study presented in this paper we use the 3-component analysis of images taken with the Extreme UV Imaging Telescope (\cite[{Delaboudini{\`e}re} {et~al.} 1995]{EIT1995}, EIT) onboard SOHO carried out by \cite[{Barra} {et~al.} (2009)]{Barra2009}. The segmentation analysis is based on the so-called {\textit{fuzzy clustering}}, a robust and fast technique that allows to monitor and track active regions on the solar disk. The analysis provides the disk-integrated relative area coverage of coronal holes, the quiet corona, and active coronal regions detected on the solar disk. However, extended active regions located beyond the solar limb are not yet taken into account. To compensate the missing radiation, we assume here an additional 20\% increase of the irradiance due to the flux coming from the extended corona. The detailed feature analysis of the off-limb features will be included in future work.

\begin{figure}[tt]
\vspace{1cm}
\centering
\includegraphics[width=.9\linewidth]{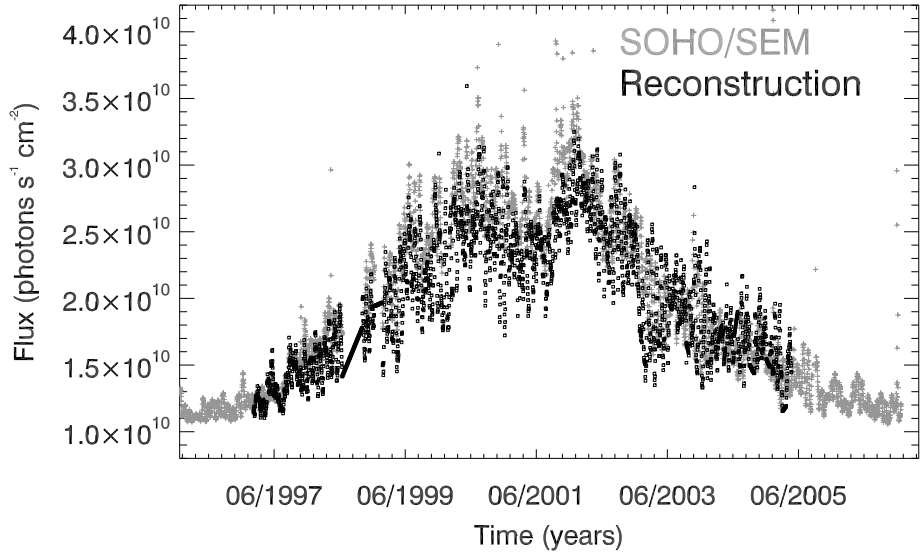}
\caption{Preliminary results of the reconstruction of the EUV photon flux at 1 AU for the wavelength range from 26 to 34 nm (black squares) compared with the SOHO/SEM observations for the same wavelength range (grey crosses). \label{fig:EUV_reconstr}} 
\end{figure}
\section{Results}\label{sec:results}
Fig.\,\ref{fig:EUV_reconstr} shows the reconstruction of the EUV from 26 to 34\,nm from Feb 14, 1997 to May 1, 2005 compated with the SOHO/SEM observations covering the same wavelength range. The average quiet corona is represented as 75\% of the quiet corona, 22\% of the coronal network, and 3\% of the active coronal network. This combination has already been used by \cite[Haberreiter (2011)]{HaberreiterSoPh2011} for modeling the EUV for solar minimum conditions. For the coronal holes we use a contrast of 0.5 with respect to the quiet Sun intensity. We tested the effect of the contrast of coronal holes and conclude that the contrast of the coronal holes shows a minor effect on the overall EUV variability. The variations in the EUV are mainly driven by the varying area coverage of the bright active regions. 

As the segmentation analysis of \cite[{Barra} {et~al.} (2009)]{Barra2009} includes only on-disk features, the missing radiation from the extended corona has to be compensated by assuming that 20\% of the on-disk radiation account for the radiation of the extended corona. The off-limb contribution will be studied in detail in the near future. A refinement of the features in the segmentation analysis will also be addressed in upcoming studies.  Finally, for longer wavelength ranges we also need to account for the detailed center-to-limb variations of the contribution from the chromospheric features. Here, the goal is to implement PSPT images (see e.g. \cite[Ermolli {et~al.} 2007]{Ermolli2007AA}). Given the limitations of these preliminary results the agreement with the SOHO/SEM observations is very promising. 

\section{Conclusion}

We have presented a preliminary version of the reconstruction of the EUV variations from Feb 14, 1997 to May 1, 2005 for the wavelength range between 26 and 34\,nm and compared it with observations taken with the CELIAS/SEM instrument. This work is based on the calculation of spectra in the EUV with the SolMod3D code and the filling factors for three coronal components. The good agreement of the reconstruction with the observed irradiance variations shows that our approach is suitable for the study of the EUV variability. Nevertheless, further analysis is required, in particular the study of the off-limb contribution to the EUV irradiance and the consideration of additional coronal activity features.

\section*{Acknowledgements}
We thank Veronique Delouille and Cis Verbeeck for kindly providing the segmentation analysis of the EIT data. SOHO is a project of international cooperation between ESA and NASA. MH acknowledges support by the Holcim Foundation for the Advancement of Scientific Research. 
%/////////////////////////////////////////////////////////////////////////////
%\bibliography{ref_all06}

\begin{thebibliography}{21}
\expandafter\ifx\csname natexlab\endcsname\relax\def\natexlab#1{#1}\fi

\bibitem[{Barra} {et~al.}(2009)]{Barra2009}
{Barra}, V., {Delouille}, V., {Kretzschmar}, M., \& {Hochedez}, J. 2009, {\textit{A\&A}}, 505, 361

\bibitem[{Delaboudini{\`e}re} {et~al.}(1995)]{EIT1995}
{Delaboudini{\`e}re}, J.-P., {Artzner}, G.~E., {Brunaud}, J., {et~al.} 1995,
  {\textit{Solar Phys.}}, 162, 291

\bibitem[{Didkovsky} {et~al.}(2010)]{Didkovsky2010ASPC}
{Didkovsky}, L.~V., {Judge}, D.~L., {Wieman}, S.~R., \& {McMullin}, D. 2010, in: {S.~R.~Cranmer, J.~T.~Hoeksema,
  \& J.~L.~Kohl} (eds.), {\textit{SOHO-23: Understanding a Peculiar Solar Minimum}}, Astronomical Society of the Pacific Conference Series, Vol. 428, p. 73

\bibitem[{Dudok de Wit} {et~al.}(2009)]{DudokdeWit2009GRL}
{Dudok de Wit}, T., {Kretzschmar}, M., {Lilensten}, J., \& {Woods}, T. 2009, {\textit{GRL}}, 36, 10107

\bibitem[Ermolli {et~al.} (2007)]{Ermolli2007AA}
{Ermolli}, I. and {Criscuoli}, S. and {Centrone}, M. and {Giorgi}, F., \& {Penza}, V. 2007, {\textit{A\&A}}, 465, 305

\bibitem[{Fontenla} {et~al.}(2009)]{Fontenla2009ApJ}
{Fontenla}, J.~M., {Curdt}, W., {Haberreiter}, M., {Harder}, J., \& {Tian}, H.
  2009, {\textit{ApJ}}, 707, 482

\bibitem[{Haberreiter} (2011)]{HaberreiterSoPh2011}
{Haberreiter}, M. 2011, \textit{Solar Phys.}, 274, 473

\bibitem[{Haberreiter} {et~al.}(2005)]{Haberreiter2005ASpR}
{Haberreiter}, M., {Krivova}, N.~A., {Schmutz}, W., \& {Wenzler}, T. 2005, {\textit{Adv. Sp. Res}}, 35, 365

\bibitem[{Hovestadt} {et~al.}(1995)]{SEM}
{Hovestadt}, D., {Hilchenbach}, M., {B{\"u}rgi}, A., {et~al.} 1995, {\textit{Solar Phys.}}, 162, 441

\bibitem[{Kretzschmar} {et~al.}(2009)]{Kretzschmar2009AcGeo}
{Kretzschmar}, M., {Dudok de Wit}, T., {Lilensten}, J., {et~al.} 2009, {\textit{Acta
  Geophysica}}, 57, 42

\bibitem[{Lean} {et~al.} (2011)]{Lean2011JGR}
{Lean}, J.~L., {Woods}, T.~N., {Eparvier}, F.~G., {et~al.} 2011, {\textit{JGR}}, 116, 1102

\bibitem[{Shapiro} {et~al.}(2011)]{Shapiro2011AA}
{Shapiro}, A.~I. and {Schmutz}, W. and {Rozanov}, E. and {Schoell}, M. and 
	{Haberreiter}, M. and {Shapiro}, A.~V. and {Nyeki}, S. 2011, {\textit{A\&A}}, 529, A67

\bibitem[{Solomon} {et~al.} (2010)]{Solomon2010}
{Solomon}, S.~C., {Woods}, T.~N., {Didkovsky}, L.~V., {Emmert}, J.~T., \&
  {Qian}, L. 2010, {\textit{GRL}}, 37, 16103

\bibitem[{Thuillier} {et~al.}(2011)]{Thuillier2011}
{Thuillier}, G., {Claudel}, J., {Djafer}, D., {et~al.} 2011, {\textit{Solar Phys.}}, 268, 125

\bibitem[{Tobiska} {et~al.}(2000)]{Tobiska2000}
{Tobiska}, W.~K., {Woods}, T., {Eparvier}, F., {et~al.} 2000, {\textit{Journal of
  Atmospheric and Solar-Terr. Phys.}}, 62, 1233

\bibitem[{Warren}(2006)]{NRLEUV} {Warren}, H.~P. 2006, {\textit{Ad. Sp. Res}}, 37, 359

\end{thebibliography}
%\bibliographystyle{aa}

\begin{discussion}

\discuss{Linsky}{I would like to draw your attention to a recently published paper by Juan Fontenla in which he updated the atmospheric models of the solar atmosphere (Fontenla et al. 2011). In particular the formation of the Lyman continuum has been improved. These latest set of solar atmosphere models has been successfully employed to reproduce the Lyman continuum of stars with various effective temperatures (Linsky et al. 2012).}

\discuss{Haberreiter}{We thank Jeff Linsky for this comment and will follow up on the recommended publications.}
\end{discussion}

\end{document}